\documentstyle[12pt]{article}
\begin{document}
\title{Nambu-Goto string with massive ends at finite temperature}
\author{G. Lambiase\thanks{Permanent address:
Dipartimento di Fisica Teorica
e S. M. S. A., Universit\'a di Salerno,
84081 Baronissi (SA), Italia},~~~V.V. Nesterenko \\
{\small \it Bogoliubov Laboratory of Theoretical Physics}\\
{\small \it Joint Institute for Nuclear Research, Dubna, 141980,
 Russia}}
\date{}
\maketitle
\begin{abstract}
It is shown that the critical (or deconfinement) temperature for the
Nambu-Goto string connecting the point-like masses (quarks) does not
depend on the value of these masses. This temperature turns out to be
the same as that in the case of string with fixed ends (infinitely
heavy immobile quarks).
  \end{abstract}
 \newpage

\section{Introduction}

In papers~[1,2] the dependence of the interquark potential on the
quark masses has been investigated in the model of the relativistic
string with ends loaded by point-like masses. It was shown that
allowance for the finite values of quark masses leads to considerable
corrections to string potential in comparison with calculations when
the string ends are fixed (immobile quarks with infinitely heavy
masses).    The critical radius\footnote{As known~[3] the potential
generated by string is not determined for any distances between
quarks no sooner than at $R >R_c$,  where $R_c$ is the critical
radius in the Nambu-Goto string with fixed ends which is determined
by the string tension $M_0^2$:  $R_{c}=\pi\,(D  -  2)/(12\,M_0^2)$.}
in the string potential turns out to be dependent on the quark masses
especially in the case of the asymmetric configurations when string
ties together light and heavy quarks~[2].

Along with the critical radius of the interquark potential, an
important parameter of the string model of hadrons is the critical or
deconfinement temperature~[4]. This prediction of the string model is
directly compared with the lattice simulations in the framework of
the gauge theories~[5]. In this connection the investigation of the
quark mass contribution to the critical string temperature is of
certain interest. It is this problem that will be treated in this
paper.

Usually one believes that in string models the critical radius of the
potential $R_c$ and the critical temperature are related. For
example, in the Nambu-Goto string with fixed ends the relation $2\,
R_c\, T_c=1$ holds. As the critical radius $R_c$ depends on the quark
masses~[1], one could expect that this will be valid for the critical
temperature $T_c$ also. However this is not so. The deconfinement
temperature in the Nambu-Goto string with point-like masses (quarks)
at its ends turns out to be independent on the quark masses, this
temperature being the same as that in the string with fixed ends.

 The layout of the paper is the following. In Sec.~II the definition
of the critical temperature in string models is reminded in short. In
Sec.~III a new approach to calculation of the Casimir energy at
finite temperature  in the Nambu-Goto string of finite length  is
suggested.  The temperature dependence of this energy  determines the
deconfinement or critical temperature in the string model under
consideration. In Sec.~IV it is shown that the critical temperature
in the Nambu-Goto string model with fixed ends and with point-like
masses attached to string ends is the same. In Conclusions (Sec.~V)
the obtained results  are discussed in short.

\section{Critical temperature in string models}
The critical temperature (or the temperature of deconfinement) in
string models is determined in the following way. If $V(R,\,T)$ is
the string potential calculated at finite temperature $T$ then its
asymptotics at large distances is
\begin{equation} V(R,\,T) \to \sigma
(T)\cdot R, \quad R \to \infty,
\end{equation}
where $\sigma(T)$ is an effective string tension depending on the
temperature~$T$. At $T=T_c$ the string tension vanishes
\begin{equation}
\sigma(T_c) =\lim_{R\to \infty}R^{-1}V(R,T_c)=0\,{.}
 \end{equation}

The string potential at finite temperature is calculated directly in
the same way as at $T=0$~[1]. As a result one obtains the well known
square-root expression
\begin{equation}
V(R,T,m)=
M_0^2R\sqrt{1+\frac{2(D-2)}{M_0^2R}\,E_C(R,T,m)}\,{,}
\end{equation}
where $M_0^2$ is the string tension at zero temperature
$\sigma(T=0)=M_0^2$, $m$ is the quark mass, $E_C(R,\,T,\,m)$ is the
renormalized Casimir energy for one transverse degrees of freedom in
string model, and $D$ is the dimension of the space-time. Usually one
puts $D=4$. Hence, determination of the critical temperature requires
calculation of the Casimir energy at finite  temperature  in the
string model under investigation.

\section{Casimir energy at finite temperature}
\setcounter{equation}{0}
The Casimir energy in the Nambu-Goto string at finite temperature is
given by the following formula~[1]
\begin{equation}
E_C(R,T)=\frac{T}{2}\sum_{n=-\infty}^{+\infty}\sum_{k=1}^{\infty}\ln
(\Omega^2_n+\omega^2_k)\,{,}
\end{equation}
where $\Omega_n$ are the Matsubara frequencies $\Omega_n=2\pi
nT,\quad n=0,\pm 1,\ldots $; $T$ is the temperature, and $\omega_k$
are the eigenfrequencies of the string determined by the boundary
conditions for the string coordinates. If the string ends are fixed
(immobile quarks) then
\begin{equation}
\omega_k=\frac{k\pi}{R}, \quad k=1,2,\ldots \,{.}
\end{equation}
In the case of the string with masses at its ends the frequencies
$\omega_k$ are the positive roots of the following equation~[1,2]
\begin{equation}
\tan(\omega \, R)=\frac{2 \omega m}{\omega^2-M_0^2}\,{.}
\end{equation}
For simplicity the symmetric quark configuration is considered
$m_1=m_2=m$. Keeping in mind that the critical temperature is
determined by the value of $E_C(R,T)$ at $R\to \infty$ (see
Eqs.~(2.2) and (2.3)),  it is
convenient, in the case of string frequencies (3.1), to do this
limiting transition directly in Eq.~(3.1) substituting the summation
over $k$ by integration
\begin{equation}
E^{\mbox{fixed}}_C(R\to \infty , T)=
\frac{T}{4}\sum_{n=-\infty}^{+\infty}\int\limits_{-\infty}^{+\infty}dk
 \ln \left [  \Omega^2_n +\left ( \frac{ k\pi}{R}\right )^2
 \right ]\,{.}
 \end{equation}
 Analytical regularization gives the following finite value of the
 divergent integral over $dk$
 \begin{equation}
 \int\limits_{-\infty}^{+\infty}\frac{dk}{2\pi}\ln (a^2+k^2)=a\,{.}
 \end{equation}
Taking account of this one obtains from Eq.~(3.4)
\begin{equation}
E_C^{\mbox{fixed}}(R\to\infty,T)=\frac{RT}{2}\sum_{n=-\infty}^{+\infty}
\Omega_n
=2\pi R T^2\sum_{n=1}^{\infty}n=
2\pi R T^2\zeta (-1)= - \frac{\pi R T^2}{6}\,{,}
\end{equation}
where $\zeta (z)$ is the Riemann zeta function. Substituting (3.6) in
(2.3) we find the critical temperature in the Nambu-Goto string with
fixed ends
\begin{equation}
\frac{T_c}{M_0}=\sqrt {\frac{3}{\pi(D-2)}}\,{.}
\end{equation}

However this method cannot be applied to string frequencies
determined by Eq.~(3.3). Investigation of the double sum in (3.2)
without introducing    the integration over $dk$ (see [6]) is again
based on the fact that frequencies $\omega_k$ are multiples of
$\pi/R$.

Here we suggest a new method for calculating the Casimir energy at
finite temperature that equally well works both with frequencies
(3.2) and with string frequencies determined by Eq.~(3.3). The idea
of the method is the following. At first we represent the
renormalized Casimir energy at zero temperature in terms of the
integral over string eigenfrequencies. In other words, we obtain the
spectral representation for $E_C(R,T)$
\begin{equation}
E_C(R,T=0)=\int\limits_{0}^{\infty}d\omega {\cal E}_C(R,\omega)\,{,}
\end{equation}
and then pass in a standard way~[7,8] from integration to summation
over the Matsubara frequencies $\Omega_n=2 \pi n T, \quad n=0,\pm 1,
\pm 2, \ldots$.  Practically this can be done by the following
substitution in~(3.8)
\begin{equation}
d \omega\to 2\pi T\, \delta (\omega - \Omega_n)\,{,}
\end{equation}
with result
\begin{equation}
E_C(R,T)= 2 \pi T \mathop{{\sum}'}_{n=0}^{n=\infty}{\cal
E}_C(R,\Omega_n)\,{.}
\end{equation}
The prime in sum sign means that the term with $n=0$ should be taken
with multiplier~1/2.

In our preceding paper~[1] we derived the following integral
representation for the Casimir energy in the Nambu-Goto string with
fixed ends
\begin{equation}
E^{{\mbox fixed}}_C= \frac{1}{2\pi}\int\limits_{0}^{\infty}d\omega \,
\ln \left ( 1- \exp {(-2 \omega R)}
\right )= - \frac{R}{\pi}\int\limits_{0}^{\infty}
\frac{\omega\, d\omega}{\exp{(2\omega R)}-1}\,{.}
\end{equation}
The last expression in this formula is obtained by integration by
parts.  Omitting the sign minus in (3.11) we see that the spectral
density of energy in this formula is of plankian form for the
one-dimensional black body, the effective temperature of the string
vacuum being equal to $(2 R)^{-1}$.  Applying the algorithm explained
above we find
\begin{equation}
E^{{\mbox fixed}}_C(R,T)=- 4 \pi T^2 R
\mathop{{\sum}'}_{n=0}^{\infty}\frac{n}{\exp{(4 \pi n R T)}-1}\,{.}
 \end{equation}
Integration by parts in (3.11) was required for obtaining the term
with $n=0$ in the sum (3.12) without divergencies. In analogous
calculations in statistical physics [7,8] for overcoming the stated
difficulty the Casimir force is calculated at first and then on this
basis the corresponding potential is recovered.

The sum in (3.12) can be evaluated in two limiting cases, for large
and for small temperatures. At large $T$ the main contribution in
(3.12) is due to the term with $n=0$
\begin{equation}
E^{{\mbox fixed}}_C(R, T\to \infty)= - \frac{T}{2}\,{.}
\end{equation}
At small $T$ the Euler-Maclaurin  formula
\begin{equation}
\mathop{{\sum}'}_{n=0}^{\infty}f(n)= \int\limits_{0}^{\infty}f(x)
 \, dx -
\frac{1}{12} f'(0)
\end{equation}
can be used.
In the case under consideration
\begin{equation}
f(x)=\frac{x}{\exp{(4\pi T R x)}-1}\quad  \mbox{and}\quad f'(0)
=-\frac{1}{2}\,{.}
\end{equation}
As a result, we obtain for small~$T$
\begin{equation}
E^{{\mbox fixed}}_C(R,T)= -\frac{\pi}{24\, R}- \frac{\pi T^2R}{6}\,{.}
\end{equation}

Proceeding from the physical consideration, it is clear that the
string picture of quark confinement inside the hadrons is applicable
only at low temperatures. In string models the temperature scale is
determined by the string tension  $M_0 \sim 0.4$~GeV. Hence under
finding the critical temperature (or the deconfinement temperature)
in string models, the region of small  temperatures should be
considered~[4]. With allowance for all this, we have to substitute in
(2.1) and (2.3) the expression for the Casimir energy at small~$T$
(Eq.~(3.16)). After taking the limit $R\to\infty$ in (2.2)
contribution of the first term in (3.16) to the effective string
tension vanishes and we obtain the critical temperature~(3.7).

By making use of the formula (3.16) one can formally introduce a
critical temperature for string of a finite length~$\bar T_c(R)$. To
this end, $\lim _{R\to \infty}$ should be removed from the definition
(2.2). At this temperature, the effective tension of the string of
finite length $R$ vanishes. It is obvious that the string critical
temperature defined in this way will be dependent on the string
length~$R$. By making use of (3.16), (2.2), and (2.3) we obtain
\begin{equation}
\bar \tau^2_c(\rho)=\tau ^2_c-\frac{1}{4\rho^2}\,{,}
\end{equation}
where the following dimensionless variables are introduced
\begin{equation}
\tau_c=T_c/M_0,\quad \rho= M_0R\,{.}
\end{equation}

It is clear that when $\rho \to \infty $ the critical temperature
$\bar \tau_c(\rho)$ dependent on the string length $\rho$ tends to
its limiting value $\tau _c$ from below because the existence of
longer strings requires a greater work for their splitting (phase
transition) than it takes place in the case of  short strings.

In the framework of other approach, the critical temperature for the
Nambu-Goto string of finite length has been considered in paper~[9].
The results obtained there probably imply that at $\rho \to \infty$
$\tau _c(\rho)$ tends to $\tau _c$ from below.

\section{Critical temperature for the Nambu-Goto string
with massive ends}
\setcounter{equation}{0}

The method for calculating the Casimir energy at finite temperature
presented in preceding section can be directly applied to the
Nambu-Goto string with massive ends. In our preceding paper [1] the
following integral representation for the Casimir energy at zero
temperature in this string model has been derived
\begin{equation}
E_C(R,T=0,m)=\frac{1}{2\pi}\int\limits_{0}^{\infty} d\omega\ln \left [1
- e^{-2 \omega R}
\left ( \frac{\omega m -M_0^2}{\omega m +M_0^2}
\right )^2                                    \right ]\,{,}
\end{equation}
where $m$ is the quark mass. For simplicity, the quarks with equal
masses are considered. The Casimir energy and the string potential
for different quark masses $m_1\ne m_2$ have been considered in~[2].

Integrating by parts in (4.1) one obtains
\begin{equation}
E_C(R, T=0,m)==- \frac{R}{\pi}\int\limits_{0}^{\infty}
\frac{\omega\,d\omega}{g(\omega)-1}\left (1-
\frac{2mM_0^2}{R(\omega^2m^2-M_0^4)}
\right )\,{,}
\end{equation}
where the notation
\begin{equation}
g(\omega)=e^{2 \omega R}\left (\frac{\omega m+M_0^2}{\omega m -
M_0^2}\right )^2
\end{equation}
is introduced for smothering. It is interesting to note that
the spectral  density  of energy in
(4.2) is not plankian now.

Along similar lines we arrive at the Casimir energy at finite
temperature
\begin{equation}
E_C(R, T,m)=-4 \pi T^2R \mathop{{\sum}'}_{n=0}^{n=\infty}
\frac{n}{g(2 \pi n T)-1}
\left ( 1-\frac{2 m M_0^2}{R(4 \pi ^2n^2T^2m^2-M_0^4)}
\right ),
\end{equation}
where $g(x)$ is defined in (4.3). When $m\to \infty$ or $m\to 0$
Eq.~(4.4) reduces to (3.12).

It is important to note that the Casimir energy at finite temperature
given by the sum (3.12) is free of any   divergencies. It is a direct
consequence of using, under derivation of (3.12), the integral
representation (4.1) for the {\it renormalized} Casimir energy at
zero temperature.

In the limit $T\to \infty$ the Casimir energy (4.4) tends to the
value
\begin{equation}
E_C(R,T\to \infty,m) = - \frac{T}{2}\left (1+\frac{2m}{M_0^2 R}
\right )\,{.}
\end{equation}
At low temperature the estimation of the sum in (4.4) can be done by
making use of the Euler-Maclaurin formula (3.14) again. In order to
find the second term in the right-hand side of (3.14) the quantity
$F'(0)$ should be calculated, where
\begin{equation}
F(x)=\frac{x(1-\eta x)^2}{e^{2Rx}(1+\eta x)^2-(1-\eta x)^2} \left (
1-\frac{2 \eta}{R(\eta^2 x^2-1)}
\right )\,{.}
\end{equation}
Here $\eta$ is the ratio $m/M_0^2$. Expansion of $F(x)$ in series gives
\begin{equation}
F(x)\simeq \frac{1}{2(R+2 \eta)}\left [1- (R+2 \eta)x +{\cal O}(x^2)
\right ]\left [1 +\frac{2 \eta}{R}(1+\eta ^2 x^2)+{\cal O}(x^4)
\right ]\,{.}
\end{equation}
>From here it follows that
\begin{equation}
F'(0)=-\frac{1}{2}\left (1+\frac{2 \eta}{R}\right ) \to -
\frac{1}{2} \quad  \mbox{when}
\quad R\to \infty \,{.}
\end{equation}
In the limit $R\to \infty$ the integral term in the right-hand side
of (3.14) vanishes. Hence, the critical temperature in the model
under investigation turns out to be the same as that in the
Nambu-Goto string with fixed ends
\begin{equation}
\tau^2_c=\left (\frac{T_c}{M_0}
\right)^2=\frac{3}{\pi (D-2)}\,{.}
\end{equation}

If we again introduce into consideration the critical temperature
dependent on the string length (see the end of the preceding Section)
then this quantity proves to be dependent on the quark mass. Indeed,
now we have, instead of (3.17),
\begin{equation}
\tau^2_c(\rho,\mu)=\left (1+\frac{2 \mu}{\rho}
\right )^{-1}       \left(\tau _c^2-\frac{I(q)}{4\rho^2}
\right )\,{,}
\end{equation}
where $\mu=m/M_0,\;\rho=M_0R,\;q=M_0^2R/m=\rho/\mu$, and the function
$I(q)$ generated by the integral term in (3.14) is given by
\begin{equation}
I(q)=\frac{24}{\pi^2}\int\limits_{0}^\infty{}\frac{dz\,
z(z-q)}{e^{2z}(z+q)^2
-(z-q)^2}\frac{z^2-q^2-2q}{z+q}\,{.}
\end{equation}
When $m\to \infty$ $I(q)$ tends to 1. In order to obtain in this
limit Eq.~(3.17) from (4.10), one has to direct $\rho$ to $\infty$ in
the first multiplier in (4.10).

\section{Conclusion}
The method for calculating the Casimir energy at finite temperature
proposed here enables us to find the critical temperature in the
Nambu-Goto string with massive ends.  Beforehand it was not obvious
that this temperature proves to be the same as that in the string
model with fixed string ends. Probably it is owing to that in string
models, like in statistical models, the boundary effects appear to be
unessential for implementation of phase transitions.

 \end{document}